# Online Dynamic Parameter Estimation of an Alkaline Electrolysis System Based on Bayesian Inference

Xiaoyan Qiu
College of Electrical Engineering
Sichuan University
Chengdu, China
1425716268@qq.com

Hang Zhang
College of Electrical Engineering
Sichuan University
Chengdu, China
zhpostg@163.com

Yiwei Qiu*
College of Electrical Engineering
Sichuan University
Chengdu, China
ywqiu@scu.edu.cn

Buxiang Zhou
College of Electrical Engineering
Sichuan University
Chengdu, China
hiway_scu@126.com

Tianlei Zang
College of Electrical Engineering
Sichuan University
Chengdu, China
zangtianlei@126.com

Ruomei Qi
Department of Electrical Engineering
Tsinghua University
Beijing, China
qrm18@mails.tsinghua.edu.cn

Jin Lin
Department of Electrical Engineering
Tsinghua University
Beijing, China
linjin@tsinghua.edu.cn

Jiepeng Wang
Purification Equipment Research Institute of CSIC
Handan, China
wjpeng03@163.com

***Abstract*—When directly coupled with fluctuating energy sources such as wind and photovoltage power, the alkaline electrolysis (AEL) in a power-to-hydrogen (P2H) system is required to operate flexibly by dynamically adjusting its hydrogen production rate. The flexibility characteristics, e.g., loading range and ramping rate, of an AEL system are significantly influenced by some parameters related to the dynamic processes of the AEL system. These parameters are usually difficult to measure directly and may even change with time. To accurately evaluate the flexibility of an AEL system in online operation, this paper presents a Bayesian Inference-based Markov Chain Monte Carlo (MCMC) method to estimate these parameters. Meanwhile, posterior joint probability distributions of the estimated parameters are obtained as a byproduct, which provides valuable physical insight into the AEL systems. Experiments on a 25 kW electrolyzer validate the proposed parameter estimation method.**

***Keywords—parameter estimation, alkaline electrolysis, power to hydrogen (P2H), load flexibility, Bayesian Inference, Markov Chain Monte Carlo, polynomial approximation***

## I. Introduction

Alkaline electrolysis (AEL) is the most mature technology for large-scale industrial power-to-hydrogen (P2H) production [1]. When directly coupled with the fluctuating renewable energy sources (RESs) such as photovoltaic and wind power, the AEL system is required to operate flexibly by dynamically adjusting its hydrogen production rate [2].

The flexibility of an AEL system, characterized by loading range, ramping rate, and energy conversion efficiency, is constrained by its internal physical processes such as the polarization curve, the thermal dynamics, and the hydrogen-to-oxygen impurity crossover dynamics [2], [3]. Therefore, the flexibility of an AEL system is significantly influenced by specific parameters in these characteristics and dynamics processes. However, some of these parameters are difficult to measure directly, and some even change with time according to real-life operational experiences and our experiments. Therefore, to accurately estimate the flexibility of the AEL system, we need to estimate these related parameters in online operation accurately.

The community has been working on parameter estimation of the electrolyzers. Lebbal et al. [4] used a nonlinear least square (LS) method to identify the parameters of the electrochemical model and thermal model of the Proton Exchange Membrane (PEM) electrolyzer. Sanchez et al. [5] estimated the parameters of steady-state models of the polarization curve, the Faraday efficiency, and gas purity by nonlinear regression. Refs. [6] and [7] determined the coefficients of the PEM electrolyzer semiempirical model and graphical polarization curve using the nonlinear LS method, respectively. Particle swarm optimization (PSO) [8], [9], and the least square error method [10] were also applied to estimate the polarization curve of PEM electrolyzers.

However, these parameter estimation methods mainly focused on the polarization curve and steady-state models. Few parameter estimation methods were available for the flexibility-related dynamic processes of the AEL system. The difficulties of AEL dynamic parameter estimation include:

- The parameters needed to estimate are associated with the dynamic processes usually modeled by differential equations and may change over time in real-life operations. In this case, the estimation method needs to accommodate the dynamic processes and be computationally efficient for online operation.
- The impact of parameters on the flexibility of the AEL system could be nonlinear, and there may be strong correlations among different parameters.

To address these difficulties, this paper presents a Bayesian Inference-based Markov Chain Monte Carlo (MCMC) method to estimate the dynamic parameters of the AEL system. Further, polynomial approximation-based surrogate models are introduced to improve computational efficiency and thus enable real-time estimation. In addition to obtaining unbiased estimations of the dynamic parameters, the proposed method also obtains posterior joint probability distribution functions (PDFs) of the estimations, enabling validating the quality of the estimation and analyzing the interdependences among different parameters. Therefore, the flexibility of the P2H system can be accurately and efficiently estimated online.

This paper is organized as follows: Section II presents the process modeling of the AEL system. Section III proposes the dynamic parameter estimation method based on the Bayesian Inference and polynomial approximation. Section IV validates the proposed parameter estimation method with experiments on a 25 kW electrolyzer. Finally, Section V discusses the conclusion and future works.

## II. Dynamic Process Model of the AEL System

Fig. 1. shows the dynamic process model of an AEL system. In operation, hydrogen and oxygen gases are produced in the electrolysis stack and mixed with the lye. Then, the gas-lye mixtures arrive at separators, where gases are separated from the lye. After separation, the lye from both sides gets mixed and circulated back to the electrolysis stack.

As discussed in the Introduction, the main influential factors of the AEL system flexibility, specifically the energy conversion efficiency, loading range, and ramping rate, consist of the polarization curve, system thermal dynamics, and hydrogen-to-oxygen (HTO) gas impurity crossover. Their models are presented below.

### A. Polarization Curve Model

The energy conversion efficiency, stress on electrodes, and heat of reaction are all related to the cell voltage of the electrolyzer, which is essentially characterized by the polarization curve [1], also known as the U-I curve, as [11]:

$$U_{\mathrm{cell}} = U_{\mathrm{rev}} + rI_{\mathrm{cell}} + s\log\left[tI_{\mathrm{cell}}+1\right], \tag{1}$$

where $U_{\mathrm{cell}}$ and $I_{\mathrm{cell}}$ represent the voltage and current density of the electrolysis cell; $U_{\mathrm{rev}}$ is the reversible voltage, determined by the change of the Gibbs free energy [11] in the electrolysis reaction; parameters $s$ and $t$ in the logarithmic term determine the activation overpotential; the parameter $r$ represents the ohmic resistance, which is formulated by [5]

$$r = r_1 + r_2 T + r_3 P, \tag{2}$$

where $T$ and $P$ represent the temperature and the pressure of the stack, respectively; $r_1$, $r_2$, and $r_3$ are parameters; and $t$ is a function of the temperature $T$, as

$$t = t_1 + t_2/T + t_3/T^2, \tag{3}$$

where $t_1$, $t_2$, and $t_3$ are also parameters.

Combining (1)-(3), the polarization curve is obtained, as:

$$\begin{aligned} U_{\mathrm{cell}} &= U_{\mathrm{rev}} + \left(r_1 + r_2 T + r_3 P\right) I_{\mathrm{cell}} \\ &\quad + s\log\left[\left(t_1 + t_2/T + t_3/T^2\right) I_{\mathrm{cell}} + 1\right] \end{aligned} \tag{4}$$

The polarization curve may change over time as a result of degradation. To accurately estimate the polarization curve, the parameters, i.e., $r_1$, $r_2$, $r_3$, $s$, $t_1$, $t_2$, and $t_3$, need to be estimated in online operation.

### B. Thermal Dynamics Model

The stack temperature influences the overpotential and thus impacts the energy conversion efficiency, stress in the electrodes, system loading range, and ramping rate.

Ref. [12] proposed a lumped heat capacity model for the AEL system and is widely accepted [11], [13]. However, this model omitted the temperature differences in different components of the process shown in Fig. 1 and hence may lead to inaccuracy when the loading is fast changing. To address this, we present the following model, which consists of the thermal dynamics of the electrolysis stack, the gas-lye separators, and the heat exchangers, etc., formulated by

$$\begin{cases} C_{\mathrm{s}}\dot{T}_{\mathrm{s,out}} = \left(U_{\mathrm{cell}} - U_{\mathrm{th}}\right) I_{\mathrm{cell}}\eta_{\mathrm{F}} + U_{\mathrm{cell}} I_{\mathrm{cell}}\left(1-\eta_{\mathrm{F}}\right) \\ \qquad - \left(T_{\mathrm{s,out}} - T_{\mathrm{a}}\right)/R_{\mathrm{h,s}} - \left(T_{\mathrm{s,out}} - T_{\mathrm{s,in}}\right) C_{\mathrm{p,s}} P_{\mathrm{flow}} \\ V_{\mathrm{sep}}\rho_{\mathrm{sep}} C_{\mathrm{p,sep}}\dot{T}_{\mathrm{sep,out}} = v_{\mathrm{sep}}\rho_{\mathrm{sep}} C_{\mathrm{p,sep}}\left(T_{\mathrm{sep,in}} - T_{\mathrm{sep,out}}\right) \\ \qquad + kA_{\mathrm{c}}\left(T_{\mathrm{c,out}} - T_{\mathrm{sep,out}}\right) - \left(T_{\mathrm{sep,out}} - T_{\mathrm{am}}\right)/R_{\mathrm{h,s}} \\ V_{\mathrm{c}}\rho_{\mathrm{c}} C_{\mathrm{p,c}}\dot{T}_{\mathrm{c,out}} = v_{\mathrm{c}}\rho_{\mathrm{c}} C_{\mathrm{p,c}}\left(T_{\mathrm{c,in}} - T_{\mathrm{c,out}}\right) - kA_{\mathrm{c}}\left(T_{\mathrm{c,out}} - T_{\mathrm{sep,out}}\right) \\ T_{\mathrm{s,out}} = T_{\mathrm{sep,in}},\quad T_{\mathrm{s,in}} = T_{\mathrm{sep,out}} \end{cases} \tag{5}$$

where $C_{\mathrm{s}}$, $T_{\mathrm{s,in}}$, and $T_{\mathrm{s,out}}$ are the heat capacity, thermal resistance, inlet and outlet temperature of the stack; $U_{\mathrm{th}} = 1.48$ V is the thermal neutral voltage; $T_{\mathrm{a}}$ presents the ambient temperature; $\eta_{\mathrm{F}}$ represents the Faraday efficiency; $C_{\mathrm{p}}$ is the specific heat capacity of lye in the stack; $P_{\mathrm{flow}}$ is the mass flow of lye; $V_{\mathrm{sep}}$, $\rho_{\mathrm{sep}}$, $C_{\mathrm{p,sep}}$, $T_{\mathrm{sep,out}}$, $v_{\mathrm{sep}}$, $T_{\mathrm{sep,in}}$ are the volume, density, heat capacity, outlet temperature, flow rate and inlet temperature of lye in the separator, respectively; $k$ is the heat transfer coefficient; $A_{\mathrm{c}}$ is the heat exchanger area; $V_{\mathrm{c}}$, $\rho_{\mathrm{c}}$, $C_{\mathrm{p,c}}$, $T_{\mathrm{c,out}}$, $v_{\mathrm{c}}$, and $T_{\mathrm{c,in}}$ are the volume, density, heat capacity, outlet temperature, volume flow, and inlet temperature of cool water in the heat exchanger.

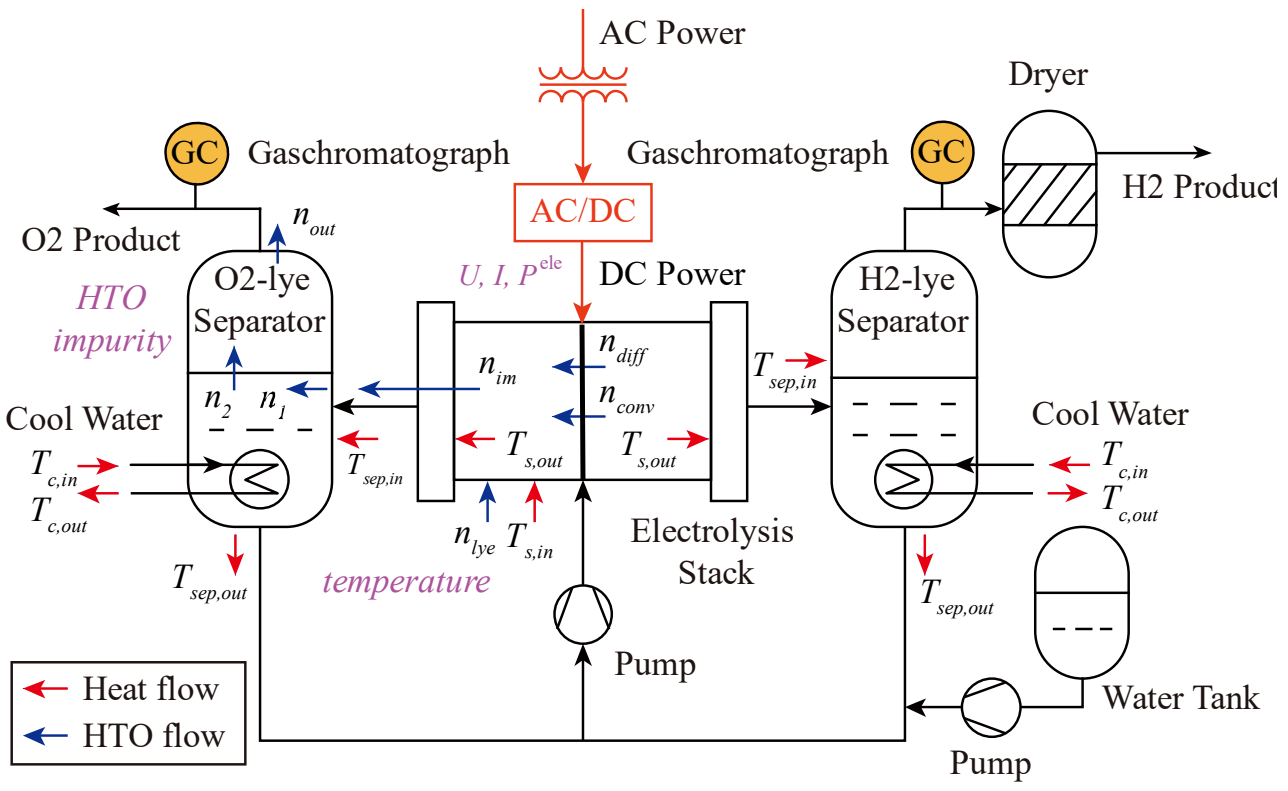

Fig. 1. Diagram of the dynamic process model of an AEL system

The main impacting parameters in the thermal dynamics model are the heat transfer coefficient $k$ of the heat exchanger, the thermal resistance $R_{\mathrm{h,s}}$ of heat dissipation, and the stack heat capacity $C_{\mathrm{s}}$, which may also vary indefinitely and need to be estimated in online operation.

### C. Gas Impurity Crossover Dynamics Model

The hydrogen impurity in the product oxygen, known as the hydrogen-to-oxygen (HTO) crossover, determines the lower limit of the loading range for an AEL [3], [14]. To avoid an explosive gas mixture, if the impurity exceeds a certain percentage (generally 2%), the AEL system shuts off. Thus, the HTO crossover dynamic process determines the lower loading limit of an alkaline electrolyzer.

The HTO impurity percentage is defined by

$$HTO = n_{\mathrm{out}}^{\mathrm{H_2}} / n_{\mathrm{pro}}^{\mathrm{O_2}}, \tag{6}$$

where $n_{\mathrm{pro}}^{\mathrm{O_2}}$ and $n_{\mathrm{out}}^{\mathrm{H_2}}$ are the molar production of $O_2$ and the $H_2$ molar impurity in the gas phase of the $O_2$-lye separator.

The HTO impurity is the result of dynamic accumulation. First, the impurity of hydrogen crossovers to the anode half-cell via diffusion and convection through the diaphragm, as well as via the lye circulation process [15].

The diffusion molar flow rate is given by Fick's law [15]:

$$n_{\mathrm{diff}}^{\mathrm{H_2}} = D_{\mathrm{eff}}^{\mathrm{H_2}} \Delta c^{\mathrm{H_2}} / \delta \approx D_{\mathrm{eff}}^{\mathrm{H_2}} S^{\mathrm{H_2}} P / \delta, \tag{7}$$

where $D_{\text{eff}}^{\text{H}_2}$ is the effective diffusion coefficient; $\Delta c^{\text{H}_2}$ denotes the differential concentration of hydrogen between the cathode and anode sides; $S^{\text{H}_2}$ is the solubility of hydrogen in the lye; $P$ is the pressure; and $\delta$ is the thickness of the diaphragm.

The convection molar flow rate follows Darcy's law [16]:

$$n_{\text{conv}}^{\text{H}_2} = \left(K/\mu\right) S^{\text{H}_2} P\left(\Delta P/\delta\right) \tag{8}$$

where $K$ represents the permeability of hydrogen through the diaphragm; $\Delta P$ is the pressure differential between the anode and cathode half-cells; $\mu$ is the dynamic viscosity of lye.

Then, the HTO impurity accumulates through three stages, i.e., the anode half-cell, the liquid phase of the separator, and the gas phase of the separator, as shown in Fig. 1. The dynamic model is established based on molar volume balance, as

$$\begin{cases} \dot{N}_{\text{an}}^{\text{H}_2} = n_{\text{im}}^{\text{H}_2} - n_1^{\text{H}_2} \\ \dot{N}_{\text{sep,liq}}^{\text{H}_2} = n_1^{\text{H}_2} - n_2^{\text{H}_2} \\ \dot{N}_{\text{sep,gas}}^{\text{H}_2} = n_2^{\text{H}_2} - n_{\text{out}}^{\text{H}_2} \end{cases}, \tag{9}$$

where $\dot{N}_{\text{an}}^{\text{H}_2}$, $\dot{N}_{\text{sep,liq}}^{\text{H}_2}$, and $\dot{N}_{\text{sep,gas}}^{\text{H}_2}$ represent the changing rates of molar quantities at the three stages, respectively; $n_{\text{im}}^{\text{H}_2}$, $n_1^{\text{H}_2}$, $n_2^{\text{H}_2}$, and $n_{\text{out}}^{\text{H}_2}$ are the molar flow rates of hydrogen, as illustrated in Fig.1, which satisfy

$$\begin{cases} n_1^{\text{H}_2} = N_{\text{an}}^{\text{H}_2} v_{\text{lye}} \Big/ \left(2V_{\text{an}}^{\text{lye}}\right) \\ n_2^{\text{H}_2} = N_{\text{sep,liq}}^{\text{H}_2} \Big/ \tau_{\text{sep}} \\ n_{\text{out}}^{\text{H}_2} = N_{\text{sep,gas}}^{\text{H}_2} n_{\text{pro}}^{\text{O}_2} \Big/ N_{\text{sep,gas}}^{\text{O}_2} \end{cases}, \tag{10}$$

where $v_{\text{lye}}$ is the flow rate of lye; $V_{\text{an}}^{\text{lye}}$ is the anode-side lye volume; $\tau_{\text{sep}}$ is the time constant of separating gas from lye.

Using (9) and (10), we have the transfer function of HTO impurity crossover, as

$$HTO(s) = \frac{n_{\text{out}}^{\text{H}_2}}{n_{\text{pro}}^{\text{O}_2}} = \frac{n_{\text{im}}^{\text{H}_2} / n_{\text{pro}}^{\text{O}_2}}{s\left(s + 1/\tau_1\right)\left(s + 1/\tau_2\right)\left(s + 1/\tau_3\right)}, \tag{11}$$

where the time constants $\tau_1$, $\tau_2$, and $\tau_3$ are given by:

$$\begin{cases} \tau_1 = 2V_{\text{an}}^{\text{lye}} \Big/ v_{\text{lye}} = V_{\text{an}}^{\text{lye}} \Big/ \left[\left(1+\phi_{\text{s}}\right) v_{\text{lye}}\right] \\ \tau_2 = N_{\text{sep,liq}}^{\text{H}_2} \Big/ \tau_{\text{sep}} \\ \tau_3 = N_{\text{sep,gas}}^{\text{O}_2} \Big/ n_{\text{pro}}^{\text{O}_2} = PV_{\text{sep,gas}} \Big/ \left(RTn_{\text{pro}}^{\text{O}_2}\right) \end{cases}, \tag{12}$$

where $R$ is the molar gas constant; the gas-lye ratio $\phi_{\text{s}}$ is a function of $n_{\text{pro}}^{\text{O}_2}$ and essentially is a function of the electrolytic current $I_{\text{cell}}$, as

$$\begin{cases} \phi_{\text{an}} = n_{\text{pro}}^{\text{o}_2} \Big/ \left(PRTv_{\text{lye}}\right) \\ n_{\text{pro}}^{\text{O}_2} = I_{\text{cell}} A_{\text{cell}} N_{\text{cell}} \Big/ \left(4F\right) \end{cases}, \tag{13}$$

where $A_{\text{cell}}$ represents the reaction area of the stack; $N_{\text{cell}}$ denotes the number of cells; $F$ is the Faraday constant.

As can be seen from the above gas impurity model, the HTO is mainly influenced by the solubility $S^{\text{H}_2}$, lye volume $V_{\text{an}}^{\text{lye}}$, and lye flow rate $v_{\text{lye}}$. However, these parameters may vary in operation and are difficult to measure directly. To address this, we need to estimate them online.

## III. Bayesian Inference-Based Parameter Estimation

### *A. Basic Theory of the Bayesian Inference*

The key flexibility-impacting parameters of an AEL system discussed in Section II are not directly measurable. Instead, we need to estimate them using available observations like the voltage, temperature, HTO impurity, etc. To achieve this and overcome the difficulties of dynamic parameter estimation discussed in the Introduction, the idea of the Bayesian Inference [17], [18] is adopted.

We suppose the relation between the parameter vector $m$ and the observation vector $d$ can be formulated by

$$d = f\left(m\right) + e \tag{14}$$

where $f\left(\cdot\right)$ is the system response function model, i.e., the differential equation models presented in Section II; and $e$ represents the vector of measurement error.

In the context of the Bayesian Inference, the parameter $m$, observation $d$, and measurement error $e$ are assumed to be random variables. The joint probability density function (PDF) of a prior estimation and the measurement error is given by

$$\pi_{prior}\left(\text{m}\right) = \prod\nolimits_{i=1}^{N_m} \pi_i\left(m_i\right), \quad \pi_{\text{e}} = \prod\nolimits_{i=1}^{N_e} \pi_{e_i}\left(e_i\right) \tag{15}$$

where $N_m$ and $N_d$ represents the number of the parameters and observations; $\pi_i\left(m_i\right)$ and $\pi_{e_i}\left(e_i\right)$ are the prior PDFs of the observation $m_i$ and error $e_i$, respectively.

The basic idea of Bayesian Inference [17] is expressed by:

$$\pi_{post}\left(m\middle|d\right) \propto \pi_{like}\left(d\middle|m\right)\pi_{prior}\left(m\right), \tag{16}$$

where $\pi_{post}\left(m\middle|d\right)$ is the posterior joint PDF of the estimated parameters; and $\pi_{like}\left(d\middle|m\right)$ is the likelihood function of the observations, as

$$\pi_{like}\left(d\middle|m\right) = \prod\nolimits_{i=1}^{N_d} \pi_{e_i}\left(d_i - f_i\left(m\right)\right), \tag{17}$$

where $\pi_{e_i}\left(\cdot\right)$ is the likelihood function of the $i$ th observation.

Substituting (17) into (16) and taking logarithm yield

$$\log \pi_{\text{post}}\left(m\middle|d\right) \propto \prod_{i=1}^{N_e} \log \pi_{e_i}\left(d_i - f_i\left(m\right)\right) + \prod_{i=1}^{N_m} \log \pi_i\left(m_i\right). \tag{18}$$

By maximizing (18), we can obtain the posterior PDF of the estimated parameters in the sense of Bayesian Inference.

### *B. The Markov Chain Monte Carlo Method*

To realize the Bayesian Inference-based dynamic parameter estimation and to accommodate the highly nonlinear nature of the dynamic processes of the AEL system, the following Markov Chain [17], [18] is first structured, as

$$m_{k+1} = m_k + \frac{\varepsilon^2}{2}\nabla \log\left(\pi_{\text{post}}\left(m_k\middle|d\right)\right) + \varepsilon\xi, \tag{19}$$

where $\varepsilon$ is the step length; $\xi$ is a vector of random variables independent of the parameters.

The next estimation $m_{k+1}$ of the parameters generated by (19) is related only to the current counterpart $m_k$ and its posterior PDF, which is known as the Markov property.

To ensure an unbiased convergence of the Markov Chain, the Metropolis-Hastings mechanism [17] is introduced. For every step, calculate

$$\alpha\left(m_k, m_{k+1}\right) = \min\left\{1, \pi_{\text{post}}\left(m_{k+1}\middle|d\right) \Big/ \pi_{\text{post}}\left(m_k\middle|d\right)\right\}, \tag{20}$$

and then draw $u : U[0,1)$. If $\alpha < u$, then we accept the new estimation $m_{k+1}$. Elsewise, we reject it and make $m_{k+1} = m_k$.

As the Markov Chain (19) reaches stationary, we obtain estimations of the parameters in the form of their posterior PDFs. Then, we can calculate the expectation of the marginal PDF of each parameter to obtain a real-valued and unbiased estimation.

**Algorithm 1** Procedure of the Proposed Dynamic Parameter Estimation Method for the Alkaline Electrolysis System

**Require:** Dynamic models of the AEL system given in Section II, initial guess of parameters $m_0$, prior PDF of parameters $\pi_{\text{prior}}(\cdot)$, step size of MCMC $N^{\text{MC}}$, the observation vector $d$

1. Use the Smolyak adaptive sparse algorithm [19] to find a polynomial surrogate model $f^*(\cdot)$ of the original dynamic model
2. Compute the initial likehood $\pi_{\text{prior}}(m_0)$
3. **for** $k = 0, \ldots, N^{\text{MC}}$ **do**
4. Generate a new sample $m_{k+1}$ in the Markov Chain by (19) with $f(\cdot)$ replaced by the polynomial surrogate model $f^*(\cdot)$
5. Compute $\alpha$ by (20)
6. Draw $u \sim U[0,1)$
7. **if** $u < \alpha$ **then**
8. set $m_{k+1} = m_{k+1}$
9. **else**
10. set $m_{k+1} = m_k$
11. **end if**
12. **end for**
13. output the mean value of $m_k$ as the estimated parameters

### C. Polynomial Approximation-Based Surrogate Model

In every step of the regular MCMC approach, the system response function $f(m)$ in (17) must be evaluated through dynamic simulation of the differential equations introduced in Section II, which is very time-consuming. To accelerate the computational speed to enable online parameter estimation, we introduce a polynomial approximation-based surrogate model in place of solving the differential equations.

The surrogate model of the dynamic response function of the AEL system is constructed as a linear combination of the orthogonal Legendre polynomials [19], written as

$$f(m) \approx f^*(m) = \sum_{i \in \mathcal{I}} c_i \Psi_i(m) \tag{21}$$

where $\Psi_i(m)$ is the multi-variate Legendre polynomial basis; $\mathcal{I}$ is the index set of the basis; the coefficients $c_i$ are calculated by the renowned collocation method, as

$$\begin{bmatrix} c_0 \\ \vdots \\ c_n \end{bmatrix} = \begin{bmatrix} \Psi_0(\hat{m}_0) & \cdots & \Psi_n(\hat{m}_0) \\ \vdots & \ddots & \vdots \\ \Psi_0(\hat{m}_n) & \cdots & \Psi_n(\hat{m}_n) \end{bmatrix}^{-1} \begin{bmatrix} f(\hat{m}_0) \\ \vdots \\ f(\hat{m}_n) \end{bmatrix}, \tag{22}$$

where $\hat{m}_i$ are the optimal sampling points, which satisfy $\Psi_{n+1}(\hat{m}_i) = 0$ [19]; $f(\hat{m}_i)$ is the corresponding system dynamic response with $\hat{m}_i$ as the parameters. For computational efficiency in finding the polynomial surrogate model, the adaptive sparse Smolyak grid algorithm is used to determine the optimal samplings points; see the detailed procedure in our previous works [20], [21].

### D. Overall Parameter Estimation Procedure

The procedure of the proposed parameter estimation method for the AEL system is summarized as Algorithm 1.

## IV. Experimental Validation

### A. Experiment Platform

We test the proposed parameter estimation by experiments on a CNDQ5/3.2 AEL system produced by the Purification Equipment Research Institute of China Shipbuilding Industry Corporation (CSIC), with a rated hydrogen production rate of 5 $Nm^3/h$ and a pressure of 3.2 MPa, as shown in Fig. 2. The main equipment parameters are listed in Table I.

The AEL system is set to operate at varying loading levels to simulate that it is connected to a volatile renewable power source. Meanwhile, we record the responses of the electrolysis system, including voltage, current, temperature, HTO impurity percentage, the openings of valves, etc. Due to the space limit, we do not show the detailed experiment data in this paper. Instead, they are available from the authors upon request.

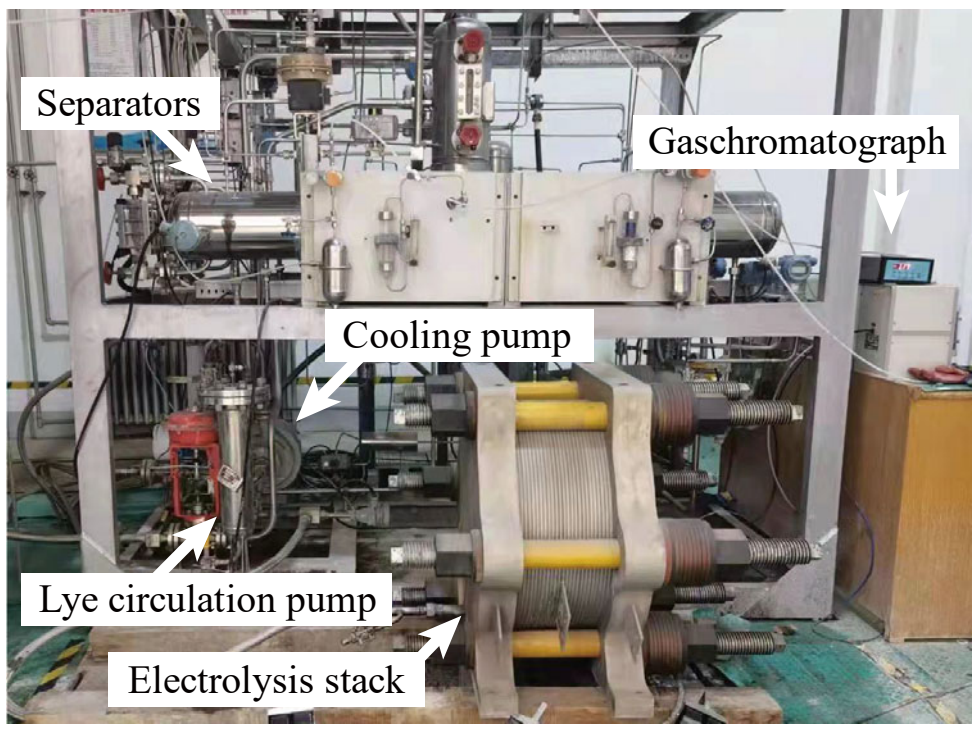

Fig. 2. The CNDQ5/3.2 alkaline electrolysis system used in the experiments

TABLE I
THE EQUIPMENT PARAMETERS OF THE EXPERIMENTAL PLATFORM

| Parameters | Value | Unit |
|---|---|---|
| ***Rated $H_2$ production rate*** | 5 | $Nm^3/h$ |
| ***Rated pressure*** | 3.2 | MPa |
| ***Operation temperature*** | 95~100 | ℃ |
| ***Rated current*** | 820 | A |
| ***Rated voltage*** | 28 | V |
| ***Number of cells*** | 26 | |
| ***Electrode area*** | 0.196 | $m^2$ |
| ***Length of separator*** | 2000 | mm |
| ***Diameter of separator*** | 219 | mm |
| ***Length of the cooling coil*** | 22 | m |
| ***Diameter of the cooling coil*** | 18 | mm |

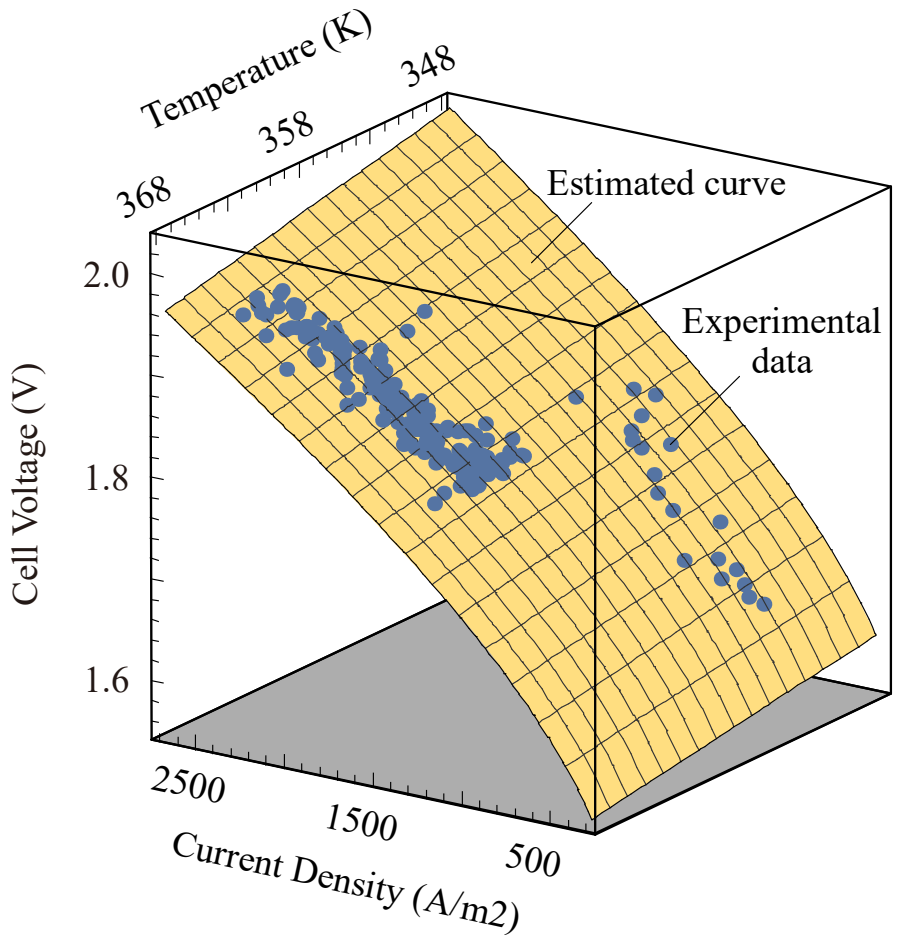

Fig. 3. The experimental data shown as the blue points and estimated polarization curve

### B. Parameter Estimation Results

The method proposed in Section III is used to estimate the key flexibility-related parameters based on the experimental data. The computation platform of parameter estimation is *Wolfram Mathematica 12.3* on a laptop with an *Intel Core i5-7200U CPU@2.50GHz*.

#### 1) Polarization Curve

First, we use the voltage, pressure, temperature, and current data (shown in the blue points in Fig. 3) of a startup process to estimate the parameters of the polarization curve discussed in Section II.A. The obtained 1- and 2-dimensional posterior marginal PDFs of the parameters are shown in Fig. 4.

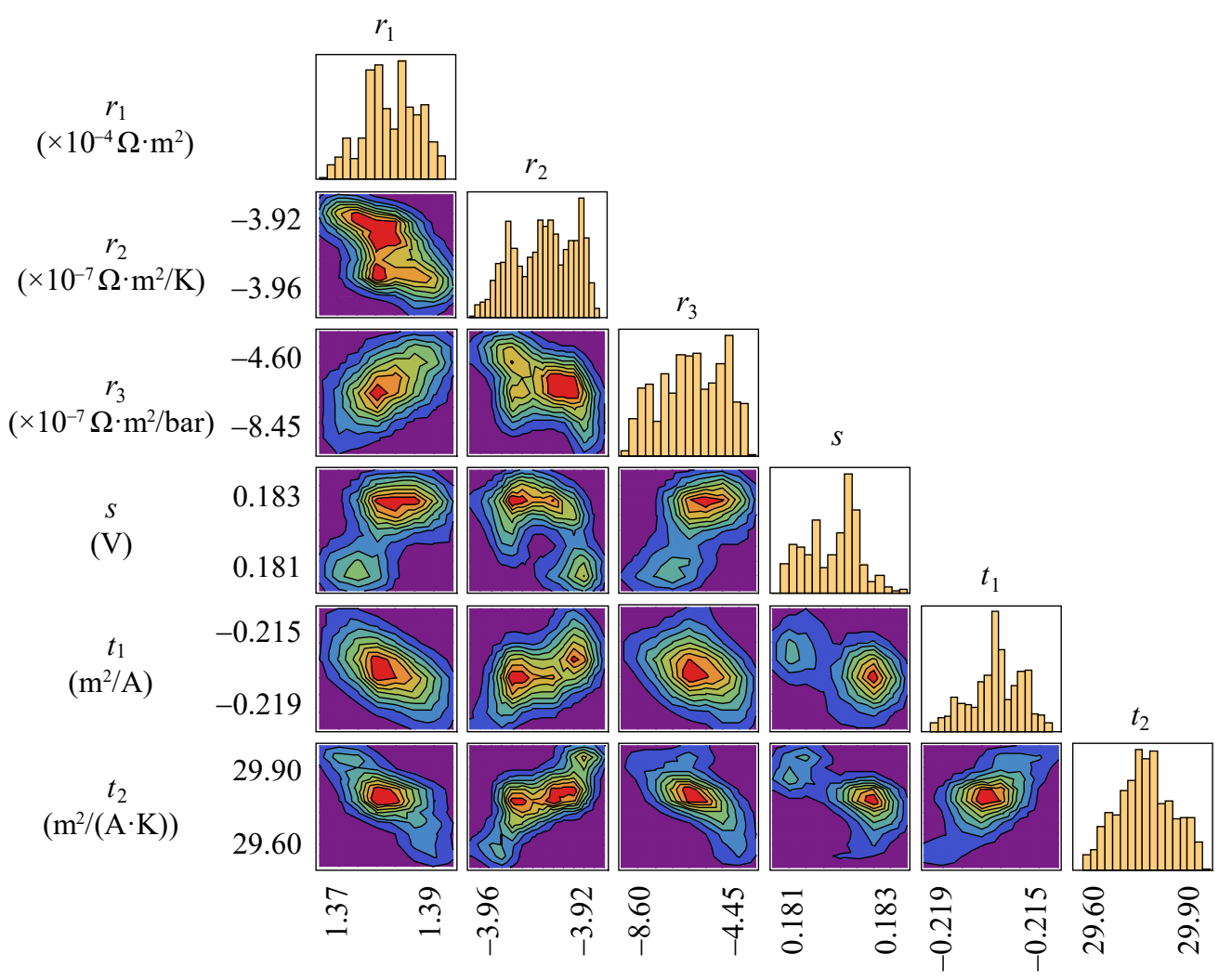

Fig. 4. The marginal and 2-dimensional PDFs of polarization curve parameter estimation obtained by the proposed method

TABLE II
THE ESTIMATED PARAMETERS OF THE AEL SYSTEM OBTAINED BY THE PROPOSED METHOD

| Parameter | Estimation | Unit |
|---|---|---|
| $r_1$ | $1.377\times10^{-4}$ | $\Omega\cdot$m$^2$ |
| $r_2$ | $-3.980\times10^{-7}$ | $\Omega\cdot$m$^2$/K |
| $r_3$ | $8.572\times10^{-7}$ | $\Omega\cdot$m$^2$/bar |
| $s$ | 0.1817 | V |
| $t_1$ | −0.2181 | m$^2$/A |
| $t_2$ | 29.90 | m$^2$/(A·K) |
| $C_s$ | $2.802\times10^{5}$ | J/K |
| $R_{h,s}$ | 0.08487 | K/W |
| $k$ | 1237.1 | W/(K·m$^2$) |
| $S^{H_2}$ | $1.025\times10^{-4}$ | mol/(L·bar) |
| $V_{an}^{lye}$ | 7.835 | L/min |
| $v_{lye}$ | 5.145 | L |

We use the expectations of the 1-dimensional posterior marginal PDFs as the estimation of the parameters, as given in Table II. The polarization curve with the estimated parameters is compared to the experimental data in Fig. 3. Quantitatively, the root mean square error (RMSE) of the estimated polarization curve is 0.00823 V, which shows that the proposed method yields a very good accuracy.

We also compare the proposed method to the widely used least square (LS) method. The RMSE obtained by the LS method is 0.00662 V. As can be seen, the RMSE of the proposed method is slightly larger than the LS method, but the difference is negligible as the parameter value is 3-orders of magnitude larger than the error. Nevertheless, the traditional LS method only provides a value of the parameter, but the proposed method is able to obtain the posterior joint PDFs of the estimated parameters. These PDFs enable us to evaluate the quality of estimation, as well as provide us insight into the nonlinear and correlated relation between different parameters, as illustrated in Fig. 4.

*2) Thermal Dynamics Model*

Then, we substitute the estimated polarization curve into the thermal dynamics model (5), and use the proposed method to estimate its parameters, namely the stack heat capacity $C_s$, the thermal resistance of heat dissipation $R_{h,s}$, and the heat transfer coefficient $k$ of heat exchanger. The estimated posterior PDFs of the parameters are plotted in Fig. 5. The expectations of the marginal PDFs are used as the estimation of the parameters, which are also listed in Table II.

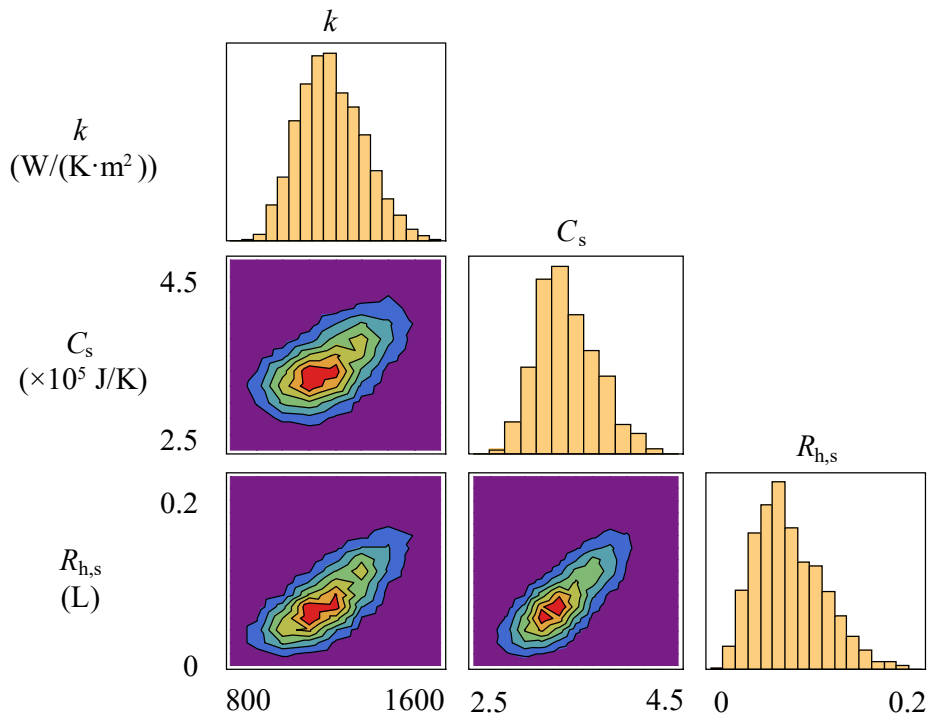

Fig. 5. The marginal and 2-dimensional PDFs of thermal dynamic model parameter estimation obtained by the proposed method

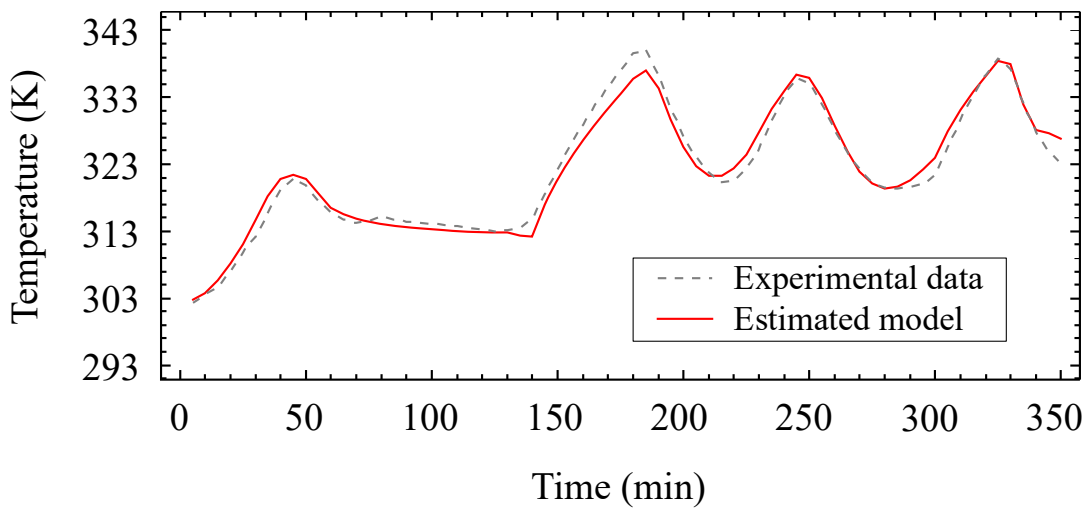

Fig. 6. The experimental temperure data and the simulation result of the dynamics model with estimated parameters

With the estimated parameters, the simulation result of the dynamic thermal model (5) is plotted in Fig. 6 in comparison with the experimental data. We can see that the estimated model again fits the actual data very well.

From the 2-dimensional marginal illustrated PDFs in Fig. 5, we can find that the three parameters are strongly correlated to each other. This, in theory, could make the parameter estimation problem very difficult to solve with the traditional methods. Nevertheless, the proposed method estimates these correlated parameters with very good accuracy.

Moreover, to showcase the improvement in computational efficiency by using the polynomial surrogate model (21), we compare the computing time with the traditional MCMC method, where the response function $f(\cdot)$ is evaluated by simulation of the differential equation model (5). The result shows that with $N^{MC}=100{,}000$, the traditional MCMC method needs $5.32\times10^4$ s to complete. Instead, by employing the polynomial surrogate model (21), the computation time of MCMC is only 722.3 s in addition to another 157.67 s for computing the polynomial approximation. This remarkably improves the computational efficiency and therefore enables online parameter estimation.

*3) Gas Impurity Crossover Dynamics Model*

Finally, we employ the proposed method to estimate the parameters of the HTO impurity crossover dynamics model, i.e., the solubility $S^{H_2}$, the lye volume $V_{an}^{lye}$ in the anode half-cell, and the lye flow rate $v_{lye}$. The estimated marginal PDFs of the parameters are shown in Fig. 7, and the estimated values of the parameters are listed in Table II.

We again perform a simulation of the HTO impurity crossover dynamic model (11) and compare the result with the experimental data, as shown in Fig. 8. As can be seen, the simulation result is consistent with the experimental result, thus validating of estimated parameters.

The proposed Bayesian Inference-based parameter estimation method also provides physical insights into the dynamic process of the alkaline electrolyzer. As seen from the 2-dimensional marginal PDFs between different parameters shown in Fig. 7, we can find that the lye volume $V_{an}^{lye}$ in the anode half-cell is almost independent of the hydrogen solubility $S^{H_2}$ and lye flow rate $v_{lye}$. However, there is a nonlinear interdependency between $S^{H_2}$ and $v_{lye}$, seen from the pattern of the 2-dimensional marginal PDF given in Fig. 7. Such physical insights could be helpful in designing and controlling the AEL system to improve its efficiency as well operational flexibility.

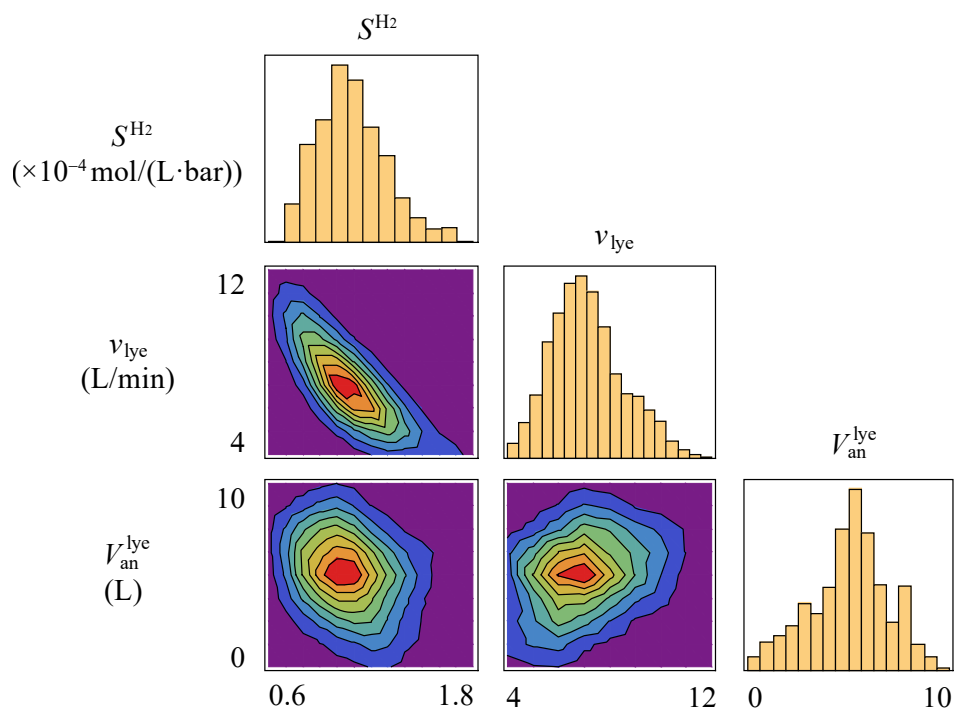

Fig. 7. The marginal and 2-dimensional PDFs of the parameter of the HTO gas impurity model, obtained by the proposed method

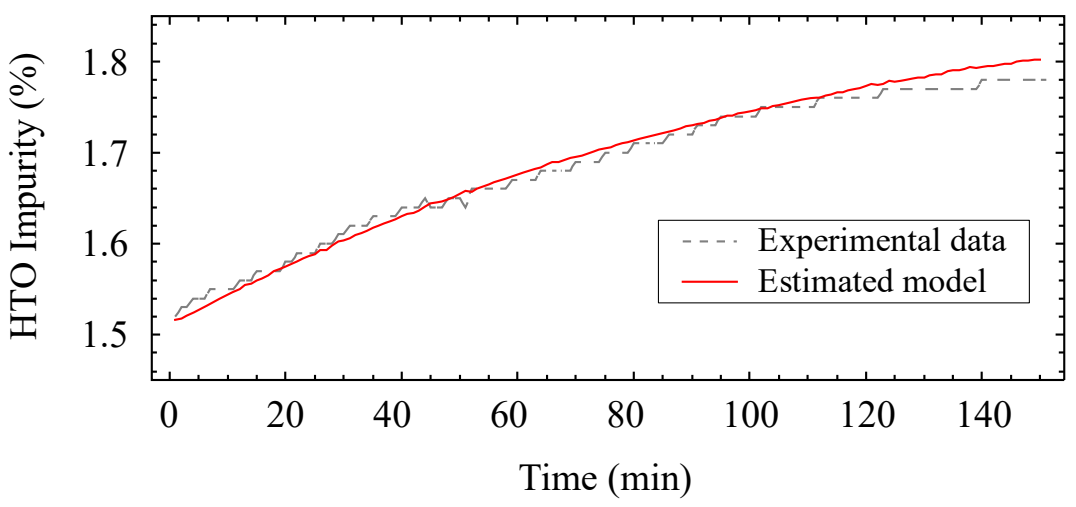

Fig. 8. The experimental HTO impurity data and the simulation result of the dynamics model with estimated parameters

## V. Conclusion

This paper presents a Bayesian Inference-based method for estimating the dynamic parameters of the alkaline electrolysis (AEL) systems. The proposed method combines the Markov Chain Monte Carlo method and the polynomial approximation technique to provide an accurate and efficient estimation of the parameters related to the operational flexibility of the AEL system. Meanwhile, joint PDFs of the estimated parameters are obtained as a byproduct, which provide valuable physical insight into the dynamics of the AEL system.

In future work, an online flexibility evaluation method for the AEL system could be developed based on the proposed parameter estimation method, which may provide information for the dispatch and control of a renewable power grid. Further, the flexibility of the AEL system could be improved by developing optimal control methods by utilizing the inter-dependences between the parameters and system flexibility.

## *Acknowledgment*

Financial supports from the National Key R&D Program of China (2021YFB4000500) and the National Natural Science Foundation of China (51907099, 51907097) are acknowledged.